# UNVEILING THE NATURE OF THE 321 S MODULATION IN RX J0806.3+1527: NEAR-SIMULTANEOUS CHANDRA AND VLT OBSERVATIONS[1]

G. L. Israel[2,3], S. Covino[4], L. Stella[2,3], C. W. Mauche[5], S. Campana[4,3], G. Marconi[6],
W. Hummel[7], S. Mereghetti[8], U. Munari[9] AND I. Negueruela[10]

2. INAF – Osservatorio Astronomico di Roma, V. Frascati 33, I–00040 Monteporzio Catone (Roma), Italy; gianluca and stella@mporzio.astro.it
3. Affiliated to the International Center for Relativistic Astrophysics
4. INAF – Osservatorio Astronomico di Brera, Via Bianchi 46, I–23807 Merate (Lc), Italy; covino and campana@merate.mi.astro.it
5. Lawrence Livermore National Laboratory, L-473, 7000 East Avenue, Livermore, CA 94550 USA; mauche@cygnus.llnl.gov
6. European Southern Observatory, Casilla 19001, Santiago, Chile; gmarconi@eso.org
7. European Southern Observatory, Karl-Schwarzschildstr. 2, D-85748 Garching, Germany; whummel@eso.org
8. CNR–IASF, Istituto di Astrofisica Spaziale e Fisica Cosmica, Sezione di Milano "G.Occhialini", Via Bassini 15, I–20133 Milano, Italy; sandro@mi.iasf.cnr.it
9. INAF - Osservatorio Astronomico di Padova, Sede di Asiago, I-36012 Asiago, Italy; ulisse@ulisse.pd.astro.it
10. Dpto. de Física, Ingeniería de Sistemas y Teoría de la Señales, Universidad de Alicante, Apdo. de Correos 99, E03080, Alicante, Spain; ignacio@astronomia.disc.ua.es

## ABSTRACT

We report on the first near-simultaneous X-ray and optical observations of RX J0806.3+1527. The source is believed to be a 321 s orbital period ultra-compact binary system hosting an X-ray emitting white dwarf. Data were obtained with *Chandra* and the ESO Very Large Telescope (VLT) in November 2001. We found an optical/X-ray phase-shift in the periodic modulation of about 0.5, strongly favoring the existance of two distinct emission regions in the two bands (for the pulsed fluxes). The *Chandra* data allow us to study, for the first time, the spectral continuum of RX J0806.3+1527 in soft X-rays. This was well fitted by a blackbody spectrum with $kT \sim 65$ eV and hydrogen column density of $N_H \sim 5 \times 10^{20}$ cm$^{-2}$. The average (unabsorbed) source 0.1–2.5 keV luminosity during the modulation-on is $L_X \sim 5 \times 10^{32}$ erg s$^{-1}$ (assuming a distance of 500 pc). Such a value is lower than the luminosity expected if stable mass transfer between two white dwarfs were driven by gravitational radiation. Evidence for absorption-like features are present in the phase-averaged spectrum at about 0.53, 0.64, and 1.26 keV, which may be attributed to heavy elements (C and N). We compare and discuss these findings with other binary systems hosting an accreting white dwarf.

*Subject headings:* stars: cataclysmic variables, white dwarfs — binaries: general, visual, close — stars: individual (RX J0806.3+1527, RX J1914.4+2456) — X-rays: binaries

## 1. INTRODUCTION

RX J0806.3+1527 was discovered in 1990 with the *ROSAT* satellite during the All-Sky Survey (RASS; Beuermann et al. 1999). However, it was only in 1999 that a periodic signal at 321 s was detected in its soft X-ray flux with the *ROSAT* HRI (Israel et al. 1999, hereafter I99; the discovery of X-ray pulsations was also reported independently by Burwitz & Reinsch 2001). Based on the large pulsed fraction ($\sim 100$%), relatively low 0.5–2.0 keV flux ($3.0-5.0 \times 10^{-12}$ erg cm$^{-2}$ s$^{-1}$), modest distance (the edge of the Galaxy is $\leq 1$ kpc in the direction of the source) and presence of a faint ($B = 20.7$) blue object in the Digitized Sky Survey 1.5″ away from the nominal X-ray position, the source was tentatively classified as a cataclysmic variable (CV) of the intermediate polar class (I99).

Subsequent deeper optical studies carried out during 1999–2001 both at the Very Large Telescope (VLT; Cerro Paranal) and at the Telescopio Nazionale Galileo (TNG; La Palma) allowed us to unambiguously identify the optical counterpart of RX J0806.3+1527, a blue $V = 21.1$ ($B = 20.7$) star consistent with that proposed by I99 and Burwitz & Reinsch (2001), and with no significant proper motion (Israel et al. 2002a, Israel et al. 2002b, hereafter I02). $B$, $V$ and $R$ time-resolved photometry revealed the presence of a $\sim 15$% modulation at the $\sim 321$ s X-ray period (Israel et al. 2002a; I02). Independently, the discovery of the optical counterpart was reported by Ramsay et al. (2002a) based on photometric observations carried out at the Nordic Optical Telescope (NOT; La Palma).

An important piece of information was based on medium-resolution spectroscopy (VLT) of this faint object (I02). The spectral study revealed a blue continuum with no intrinsic absorption lines. Broad (FWHM $\sim 1500$ km s$^{-1}$), low equivalent width ($EW \sim -2 \div -6$ Å) emission lines from the He II Pickering series (plus additional emission lines likely associated with He I, C III, N III, etc.) were instead detected. These findings, together with the period stability and absence of any additional modulation in the 1 min–5 hr period range, are interpreted in terms of a double degenerate He-rich binary (a subset of the AM CVn class; see Warner 1995 for a review) with an orbital period of 321 s, the shortest ever recorded (I02; see also Ramsay et al. 2002a). RX J0806.3+1527 was soon noticed to have optical/X-ray properties similar to those of RX J1914.4+2456, a 569 s modulated soft X-ray source recently proposed as a double degenerate system (Ramsay et al. 2000, 2002b). The study of these two objects has posed, in the last year, serious questions about their possible origin.

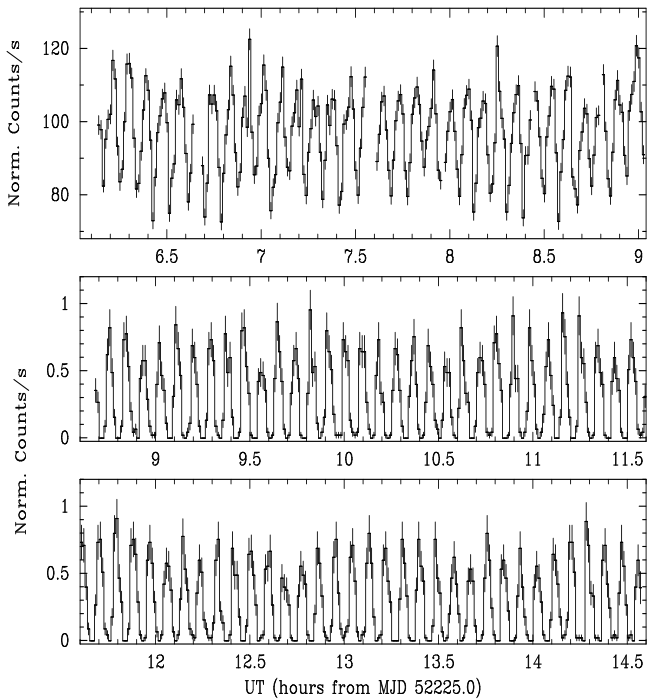

FIG. 1.— VLT FORS1 *R* special band light curve (upper panel) together with the *Chandra* ACIS-S 0.1–10.0 keV light curves (lower panels). For comparison we set the binning time to 45 s in both datasets. Time in the x-axis is hours from 2001 November 12, 0:0 UT (MJD 52225).

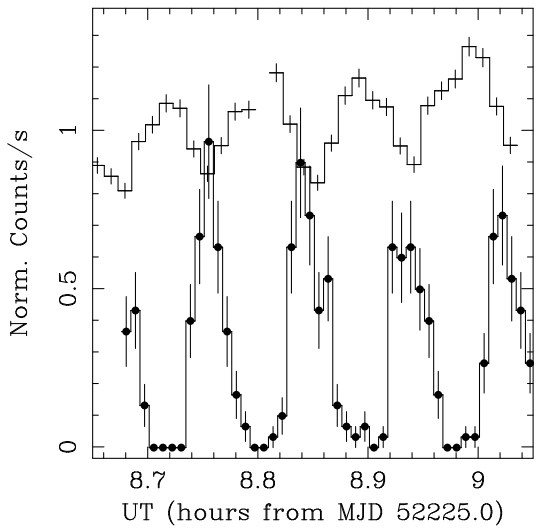

FIG. 2.— Same as Figure 1 but only for the simultaneous time interval of the VLT FORS1 and *Chandra* observations..

The nature of the soft X-ray emission detected from RX J0806.3+1527 and the twin source RX J1914.4+2456 is still debated. Several models have been proposed (see Cropper et al. 2003 and reference therein). Among these is the double degenerate binary system with mass transfer model that has been proposed in two flavors: with a magnetic (polar-like; Cropper et al. 1998) and with a non-magnetic accretor (Algol-like, also known as direct impact accretion model; Marsh & Steeghs 2002). In the latter model, and assuming a light companion ($\sim 0.13 M_\odot$), a disk would not form since the distance at which the gas stream would form the disk is smaller than the size of the accretor, resulting in the stream directly hitting the surface of the accretor. The impact region should be significantly smaller than in the magnetically-confined case, wherein the gas stream is expected to penetrate below the white dwarf photosphere, where it thermalises its energy and gives rise to the observed soft X-ray emission (Marsh & Steeghs 2002). The former scenario is similar to that of polars. However, the magnetic field of the primary does not need to be as strong as that of normal polars in order to achieve synchronism due to the small separation of the two stars. The lack of detected circular polarisation in the optical emission of RX J1914.4+2456 has been considered a possible difficulty for the polar model, unless the magnetic field strength is less than a few MG (Cropper et al. 2003). However, the absence in the *ASCA* spectrum of RX J1914.4+2456 of a hard X-ray component, as seen in many classical polars, might pose some difficulties for this scenario (however see Ramsay et al. 2002b for a possible explanation).

An alternative model involves a secondary star that does not fill its Roche lobe. As this star crosses the magnetic field of the primary an electric field (under the hypothesis of the presence of ionised material between the two stars) is induced, resulting in the heating of the polar caps of the white dwarf primary (no matter transfer is expected). This is also known as the unipolar inductor model, and is similar to the observed Jupiter-Io system (Clarke et al. 1996; Wu et al. 2002). In all the above scenarios the X-ray emitting region is thought to be self-occulted by the primary, producing the X-ray off intervals and large-amplitude modulation. Moreover, if the orbital period is indeed ultra-short, RX J0806.3+1527 and RX J1914.4+2456 are expected to be two of the strongest sources of gravitational wave radiation, easily detectable by LISA (Phinney 2003). Finally, the possibility that these two sources are stream-fed intermediate polars (IPs) seen face-on has been recently proposed under the hypothesis that the even terms of the He II Pickering series seen in RX J0806.3+1527 might be due in part to H. In the IP scenario the observed modulation would be due to the spin period of the accreting white dwarf (Norton et al. 2002). In this case, it is somewhat surprising that no evidence for the modulation at the orbital period has been found in the optical and X-ray bands.

Recently, based on optical and X-ray observations spanning over 9 years, the 321 s period of RX J0806.3+1527 was found to be decreasing at a rate of $3 \div 6 \times 10^{-11}$ s s$^{-1}$ (Hakala et al. 2003; Strohmayer 2003). This value is in agreement with the rate expected from the gravitational radiation release for two white dwarfs orbiting at 321 s. The authors finally concluded that the result favours the unipolar inductor model. Note that a similar result was obtained for RX J1914.4+2456 based on *ROSAT* and *ASCA* observations (Strohmayer 2002). In all papers the authors considered unlikely the intermediate polar scenario, although it is still marginally possible.

The lack of good quality X-ray spectra has limited the study of RX J0806.3+1527. In particular, only a very rough idea of the spectral continuum was gleaned from a $\sim 200$ s PSPC scanning observation during the *ROSAT* All-Sky Survey: for an absorbed blackbody, we inferred $kT_{bb} \sim 50^{+110}_{-35}$ eV and $N_H < 6 \times 10^{20}$ cm$^{-2}$ (see I02 for details). These data were too poor to verify whether or not the optical and X-ray flux can be accounted for with only a single spectral component. Moreover, the study of the possible presence of delays of the minima and/or maxima in the modulation shapes between the X-ray and the optical band is important for understanding the emission geometry from RX J0806.3+1527. Therefore, the optical ob-servations of RX J0806.3+1527 carried out during our *Chandra* X-ray observation allowed us to unveil some properties of this source. In this paper we report the results of the first detailed study of this object in the X-ray band. Moreover, we show the results of the phasing of the optical and X-ray light curves. Finally, we study the broad-band energy spectrum of the source and compare the observational parameters of RX J0806.3+1527 and RX J1914.4+2456 with those of a sample of CVs.

## 2. CHANDRA OBSERVATION

The field of RX J0806.3+1527 was observed with the *Chandra* Advanced CCD Imaging Spectrometer (ACIS; Garmire 1997, Bautz et al. 1998) on 2001 November 12 (from 8.6 to 14.5 UT) for an effective exposure time of 21 ks. The observation was carried out with the S3 (back-illuminated) chip with a frame read out time of 0.441 s (128 active rows). Data were reduced with CIAO version 2.2/2.3 and analysed with standard X-ray software packages (FTOOLS 5.2, XSPEC 11.2.0, XRONOS 5.19) and the latest calibration files were used in the analysis. Only one source was detected in the narrow field-of-view with position R.A. = $08^h06^m22^s.92$, DEC. = $+15°27'30''.9$ (equinox J2000; 90% confidence radius of $0''.7$), consistent with that of the optical counterpart of RX J0806.3+1527 (I99; Butwitz & Reinsch 2001), and at an average count rate of $0.29 \pm 0.01$ counts s$^{-1}$ in the 0.1–10 keV energy band. Source photon arrival times were extracted from a circular region with a radius of $2''$, including about 95% of the source photons, and corrected to the barycenter of the solar system (*axbary* in CIAO). A large-amplitude modulation was detected at a period of about 321.5 s, confirming that the source is indeed RX J0806.3+1527. Figure 1 (lower two panels) shows the *Chandra* background-subtracted 0.1-10 keV light curve of the source, where the count rates are in the $\sim 0.0$–1.0 count s$^{-1}$ range (assuming an arbitrary binning time of 45 s). No additional signal is present in the power spectrum, with a pulsed fraction upper limit of about 15–25% ($3\sigma$ confidence level) in the $\sim$1–1000 s range.

## 3. VLT OBSERVATIONS

On 2001 November 12 we obtained a Director's Discretionary Time (DDT) optical observation of RX J0806.3+1527 at the 8.2 m VLT-U4 Yepun telescope that nearly coincided with the *Chandra* pointing. The telescope was equipped with the FOcal Reducer/low dispersion Spectrograph (FORS2) and a filter referred as "*R* special"(centered at 655 nm and with FWHM = 165 nm; for comparison the *R*-Bessel filter used by ESO is centered at 657 nm with FWHM = 150 nm). Time-resolved photometry at a time resolution of 15 s was obtained for a total duration of about 3 hr (from about 6.1 to 9.0 UT) with seeing in the $0''.5$–$1''.3$ range.

The data reduction was carried out with the ESO-MIDAS (version 01SEP) system. After bias subtraction, non-uniformities were corrected using flat-fields obtained with the *R* special filter. The flux of each point source in the field-of-view was derived by means of both aperture and profile fitting photometry with the DAOPHOT II package (Stetson 1987) as implemented in MIDAS. Differential light curves were accumulated by subtracting the source magnitudes with the mean magnitude of 10 reference stars within the field-of-view (similar to the analysis reported in I02), and by correcting the center of each original bin time for the barycenter of the solar system (function *barycen* in IDL and *barycorr* in MIDAS).

Figure 1 (upper panel) shows the differential light curve of RX J0806.3+1527 rebinned to 45 s (as for the optical data in order to allow a easier comparion between the two datasets), and normalised to the mean intensity of reference star. Figure 2 shows the VLT FORS1 and *Chandra* ACIS-S light curves for the narrow time intereval during which the optical and X-ray observations were simultaneous.

### 3.1. *THE ABSOLUTE PHASE OF OPTICAL/X-RAY EMISSION*

One of the main aims of the joint *Chandra*/VLT campaign was to determine the phasing of the optical and X-ray light curves in order to get better information about the presence of one or more emission regions and their geometry. With this in mind, we first obtained an accurate measurement of the period of the 321 s modulation by using the *Chandra* observation (which is twice as long as the VLT observation): this is $321.52 \pm 0.03$ s (90% confidence level), determined by phase-folding the data. This value is consistent with the value of 321.53352(3) s inferred by Hakala et al. (2003) based on three observational runs (one of which is the optical run reported in this paper) over a baseline of about 60 days. Then, the optical and X-ray light curves were folded at 321.5332 s (the 321.53352 s period corrected for the period derivative reported by Hakala et al. 2003); the results are shown in Figure 3 (left panel). The pulsed fraction of the 321 s modulation is 100% and 14% in the X-ray and optical bands, respectively, consistent with results reported by I99, I02 and Ramsay et al. 2002a. A large phase-shift is clearly present. The folded X-ray and optical light curves have very different shapes, making difficult an unambiguous determination of their relative phase-shift. However, we note that the minimum of the optical and the maximum of the X-ray light curves are at the same phase, indicating a $\sim 0.5$ phase-shift between the optical and X-ray emission. In order to quantify the delay between the optical and X-ray band, we performed a cross-correlation function (CCF) analysis on the optical and X-ray light curves. The analysis resulted in a phase-shift of $0.48 \pm 0.03$ (90% confidence level).

Light curves were also obtained for several X-ray energy intervals: these are plotted in the right panel of Figure 3 in four different bands (0.1–0.2 keV, 0.2–0.5 keV, 0.5–10 keV and 0.1–10 keV from top to bottom). The modulation shape is highly asymmetric with the peak moving from phase 0.6 (light curve a) to phase 0.5 (light curve c). Moreover, the rising part of the modulation becomes sharper as the photon energy increases. A bump at phase 0.8–0.9 is evident at high energies (light curve c). In all cases the modulation-off count rates are consistent with zero (within the uncertainty). However, an inspection of the *Chandra* light curves shows that the count rate is consistent with the source being piled-up in the phase interval 0.5–0.9. Pile-up is an inherently nonlinear process occurring in single-photon-counting CCD cameras. Corrections are not easily applied, and a parametrization of the effects must be considered (Davis 2001). The problem is even more complex when dealing with an extremely variable source such as RX J0806.3+1527 because of the varying percentage of piled-up photons with time and/or phase. Therefore, we checked the robustness of the above results by studying the modulation shape as a function of energy for two distinct regions: the central $3 \times 3$ pixels containing nearly 90% of the source photons (where the pile-up is strong) and an annular region (between about $1''$ and $2''$ from the source position) containing nearly all the remaining source counts where the pile up is negligible. Even though the statistics of the outer annular region are poor, we note that between the two regions there might be a slightly difference in the modulation shape for energies above 0.5 keV: in the central region (right panel, curve c of Figure 3) the light curve is more asymmetric, with a steep rising part, while in the outer region the light curve is more symmetric and energy-independent (also in the outer region case the modulation-off count rates are consistent with zero). This result suggests that part of the source modulation shape variations might be ascribed to pile-up. [On the other hand, we note that the period determination is not effected by pile-up because it might modify only the shape of part of the light curve close to the modulation peak. Moreover, since the source was reasonably constant in mean flux during the observation, pile-up does not introduce a bias in the period determination.]

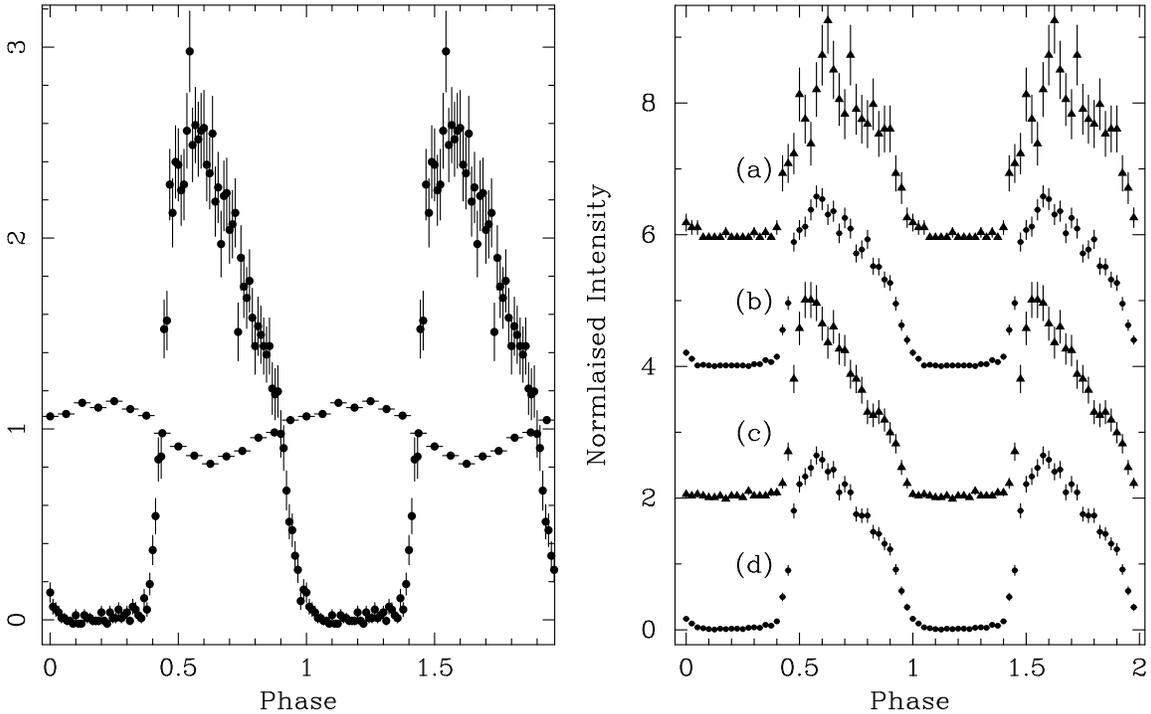

FIG. 3.— Left panel: *Chandra* folded light curve (nearly 100% pulsed fraction; eclipse is evident at phases 0.0–0.3), with superposed the VLT optical folded light curve (*R* band). Optical and X-ray peaks are phase-shifted by about 0.5. Phase 0 is arbitrary and corresponds to MHJD 52225.36153±0.00004. Righ panel: X-ray folded light curves in different energy ranges: (a) 0.1–0.2 keV, (b) 0.2–0.5 keV, (c) 0.5–10.0 keV, and (d) 0.1–10.0 keV.

### 3.2. *THE X-RAY SPECTRAL CONTINUUM*

The *Chandra* X-ray dataset was also used to carry out the first detailed spectral analysis of RX J0806.3+1527. In particular, based on the above results we checked the double degenerate and unipolar inductor model predictions, where a hot spot (a soft blackbody as a first approximation) is expected to be formed at the polar caps or at the equator of the primary star: this emission region would be responsible for irradiating part of the companion star and the accretion stream (if present). The same photons extracted for the timing analysis were used for the spectral analysis. The phase-averaged spectrum was rebinned in order to have more than 20 counts in each bin such that minimum $\chi^2$ fitting techniques could be reliably used. Background subtraction was performed using an annular region $\sim 10''$ away from the position of the source, taking care to remove the readout streak columns. Following the CIAO recipies for the ACIS-

S spectral analyses, we considered only the 0.35–10 keV energy band and took into account the known reduction in the ACIS low-energy quantum efficiency by correcting the corresponding effective area files. A first attempt at fitting the data with an absorbed blackbody model gave a reduced $\chi^2$ of 3.3 and provided evidence for the possible presence of a high energy tail (above $\sim 1$ keV), as expected for a source effected by pile-up (Davis 2001). We therefore added to the above component an ad-hoc model developed at the *Chandra* Science Data Center (see also Davis 2001) in order to take into account the effects of the nonlinear pile-up (see the SHERPA and XSPEC packages). The new phase-averaged best-fit spectrum gave a reduced $\chi^2$ of 1.13 ($\chi^2$ of 36.2 for 32 degree of freedom, hereafter d.o.f.) with parameters $kT_{bb} = 64 \pm 2$ eV and $N_H = (9.0 \pm 1.5) \times 10^{20}$ cm$^{-2}$, with a photon grade migration percentage $\alpha = 20 \pm 6\%$ (a parameter which is a measure of the pile-up); the high energy tail is no longer present. We derived an upper limit on the flux of a possible hard emission component (we assumed a power–law with slope $\Gamma \sim 1$ and kept fixed the above components) of a factor of about 100 less than the blackbody flux. The 0.1–2.5 keV unabsorbed flux is $5 \times 10^{-12}$ erg cm$^{-2}$ s$^{-1}$, corresponding to a luminosity $L_X \sim 1.5 \times 10^{32}$ erg s$^{-1}$ at 500 pc distance (all the fluxes reported in this paper superceed those reported in Israel et al. 2003). However, we note that the spectral continuum shape is continuously changing as a function of 321 s modulation phase due to the changes in the percentage of piled-up photons as a function of count rate (and, therefore, phase of the 321 s modulation).

We performed a phase-resolved spectroscopic (PRS) analysis with the above described model. The following six phase intervals were considered: 0.0–0.44, 0.44–0.54, 0.54–0.64, 0.64–0.74, 0.74–0.84 and 0.84–1.0 (see Figure 3 and 4). In all cases but in the 0.54–0.64 and 0.64–0.74 intervals we find a

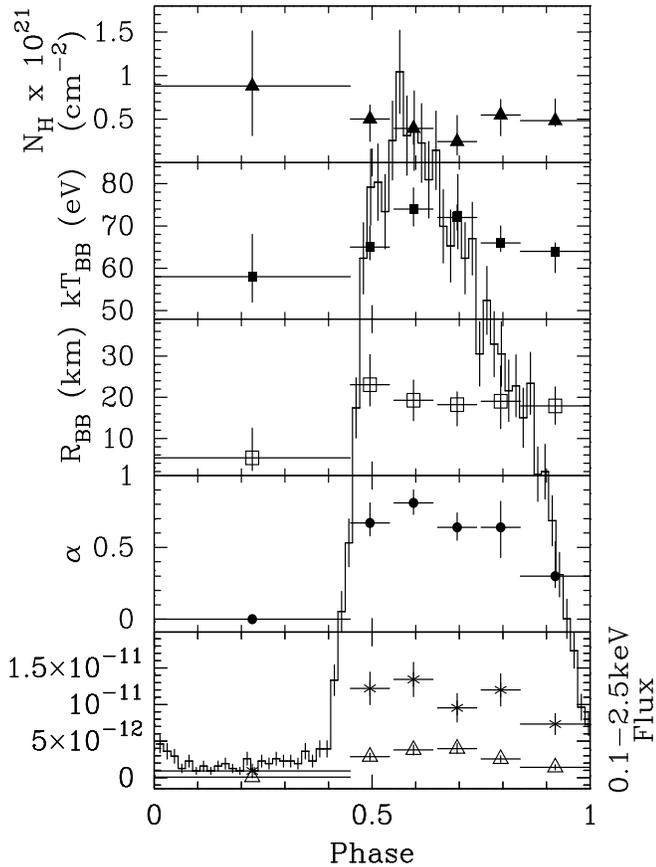

FIG. 4.— The results of the PRS analysis for the absorbed blackbody spectral parameters: absorption, blackbody temperature, blackbody radius (assuming a distance of 500 pc), photon migration grade fraction, and absorbed (triangles) and unabsorbed (asterisks) flux. Superposed is the folded X-ray light curve.

relatively good fit with a reduced $\chi^2$ in the range 0.85–1.15 (and a reduced $\chi^2$ of about 1.3 in the remaining two intervals). The results of the PRS analysis are summarised in Figure 4 where we show the phase dependence of the spectral parameters of the fits: the hydrogen column density $N_H$, the blackbody temperature $kT_{bb}$, the corresponding blackbody radius $R_{bb}$, the photon migration grade $\alpha$, and the 0.1–2.5 keV absorbed (triangles) and unabsorbed (asterisks) fluxes (from top to bottom). The error bars correspond to the 90% confidence level. We summarise the results as follows: the migration grade parameter $\alpha$ is extremely variable and confirms that pile-up played an increasingly important role as the X-ray flux approached the peak of the modulation, where up to 90% of the photons are affected. The column density is consitent with being slightly phase-dependent with $N_H \sim 5 \times 10^{20}$ cm$^{-2}$ for the modulation-on phase, and somewhat larger value for the modulation-off phase. During the modulation-on phase, the blackbody radius is consitent with being phase-independent at $R_{bb} \sim 15$–25 km (for distances of 500 pc); this represents an extremely small surface fraction of a white dwarf. Moreover, the unabsorbed 0.1–2.5 keV flux during the modulation-on phase is $\sim 1.5 \times 10^{-11}$ erg cm$^{-2}$ s$^{-1}$, corresponding to a luminosity of $\sim 5 \times 10^{32}$ erg s$^{-1}$ (assuming a distance of 500 pc, i.e., the edge of the Galaxy in the direction of the source). There is a factor of >10–20 between the on- and off-modulation fluxes (note that the phase interval 0.0-0.44 includes a small percentage of the modulation-on flux and, therefore, it is not totally representative of the modulation-off state).

An additional check was performed on a spectrum accumulated from source photons extracted from the annular region (where pile-up is negligible) used for timing analysis (see section 3.1). Even though the statistics are poor, a reduced $\chi^2$ of 0.98 ($\chi^2$ of 6.7 for 7 d.o.f.) was obtained for a best fit with the following parameters: $N_H = 9^{+8}_{-6} \times 10^{20}$ cm$^{-2}$, $kT_{bb} = 61 \pm 8$ eV, $\alpha$ consistent with zero, and a PSF-corrected unabsorbed flux of $1.5 \times 10^{-11}$ erg cm$^{-2}$ s$^{-1}$ (0.1–2.5 keV). No power-law component could be detected.

Next, we looked carefully at the phase-averaged spectrum extracted from the central region. The residuals (see bottom panel of Figure 5) show a number of absorption-like features at the following energies: $\sim 0.55$ keV, 0.65 keV, 0.8 keV and 1.2 keV. In consideration of the relatively high reduced $\chi^2$ found (1.3 and 1.1 for the phase-averaged and modulation-on spectrum, respectively), we tried to fit the features with two different models, namely an absorbed blackbody modified by absorption edges or Gaussians (leaving the width free to vary). The addition of an edge at $0.64 \pm 0.04$ keV ($\tau = 0.6 \pm 0.4$) resulted in a reduced $\chi^2$ of 1.1 (33.4/30 d.o.f.) and an F probability of 0.05 ($2\sigma$). A slightly higher significance level of $2.8\sigma$ was obtained for the inclusion of an edge at $1.26 \pm 0.05$ keV ($\tau = 3.2^{+2.5}_{-1.5}$). The addition of both edges simultaneously has a probability of $2.7\sigma$ ($\chi^2$ of 24 for 28 d.o.f.). Even though not statistically required by the data, we also added an edge at $0.53 \pm 0.04$ keV ($\tau = 0.4^{+0.2}_{-0.4}$; $\chi^2$ of 21.7 for 26 d.o.f. and probability of $1.6\sigma$). The results of this analysis are shown in Figure 5. Similar results, with comparable significance levels were obtained by using the Gaussian model. However, given the low statistical significance for the inclusion of the edges in the spectrum, we regard at these features only as a potential important result, to be confirmed by future observations.

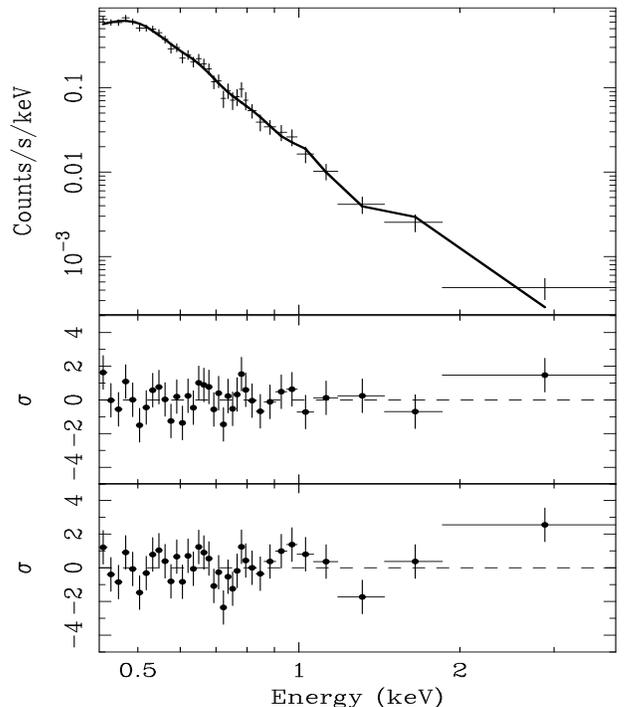

FIG. 5.— The *Chandra* ACIS-S 20 ks phase-averaged spectrum of RX J0806.3+1527 fitted by means of the results obtained in the PRS analysis and taking into account the pile-up effect. Residuals (lower panels) refer to the piled-up blackbody model (bottom panel), and the piled-up blackbody model plus the three edges described in the text (middle panel). The solid thick line in the upper panel corresponds to the latter model.

Finally, we utilized the excellent spatial resolution of the *Chandra* mirrors to performed a radial profile study of RX J0806.3+1527: using all photons but those in the 0.45–0.85 phase interval, this was found to be in good agreement with the expected point spread function up to an off-axis angle of $\sim 10''$: no extended emission component(s) have been found.

## 4. DISCUSSION

The nearly-simultaneous X-ray and optical observations of RX J0806.3+1527 carried out 2001 November with *Chandra* and VLT, respectively, provides new constraints on the nature of the source. In the following we list and discuss the results.

(i) The near 0.5 phase-shift detected between the optical ($\sim$15% modulation) and X-ray light curves strongly supports the idea that X-ray and optical photons originate from come from two different regions. Note that a similar phase-shift was measured for RX J1914.4+2456 (Ramsay et al. 2000). We also found a striking similarity between the shape of the X-ray and optical modulation of RX J0806.3+1527 and RX J1914.4+2456 (see below). X-ray photons are thought to originate from the surface of the white dwarf primary, whereas the modulated optical flux ($\sim$15% of the total) is due to the X-ray irradiated surface of the companion star (and possibly the accretion stream, if present; I02). In the Algol-like model, the X-ray modulation-off phase is easily explained by the elongated equatorial X-ray emitting region being self-eclipsed by the accreting white dwarf. In the polar-like and unipolar inductor models the X-ray emitting region (also in this case self-eclipsed) is at the magnetic polar cap(s) of the accretor.

(ii) Under the hypothesis that the blackbody provides a fair estimate of the X-ray spectral continuum of the emission region, we note that the inferred size of $\sim 20$ km is consitent with the expectations of the direct impact accretion model. Omn the contrary this value is a factor of about 10 smaller than that expected for the unipolar inductor model (Wu et al. 2002), and that measured in IPs. However, we note that the predicted radius in the unipolar inductor model depends on a number of free parameters that have not been explored yet. A more detailed study of the Io-Jupiter case may yield a reliable estimate of the expected blackbody radius for RX J0806.3+1527. On the other hand, we note that the inferred value is closer to the expectations of the direct impact accretion model.

(iii) The *Chandra* data allowed us to infer an average 0.1–2.5 keV luminosity during the on-phase of 0.2–$5 \times 10^{32}$ erg s$^{-1}$ (for a distance of 100–500 pc). In Figure 6 we plot the X-ray and optical flux measurements, together with the blackbody models obtained from the spectral fits. It is evident that a $\sim 65$ eV-temperature blackbody alone is not sufficient to account for both the optical flux and the *EUVE* upper limits (the EUVE upper limits are shown for two values of $N_H$)[1]. This is somewhat surprising since, in principle, only 15% of the total optical flux should be out of phase with respect to the X-ray emission (assuming correct the X-ray irradiation scenario). However, even though the half-cycle phase-shift we inferred between the X-ray and optical light curves argues against a single emission component producing the pulsed emission, the unpulsed optical flux (86% of the total) might be originated relatively close to the X-ray one (for example, emitted by the primary star). Moreover, the optical data is well fit by an additional blackbody model. Consequently, the inferred X-ray flux of a few times $10^{-11}$ erg cm$^{-2}$ s$^{-1}$ should be considered just a lower limit to the bolometric flux from RX J0806.3+1527. The total flux can be easily a factor of about 10 times larger than that extrapolated from the X-ray band, depending on the unabsorbed

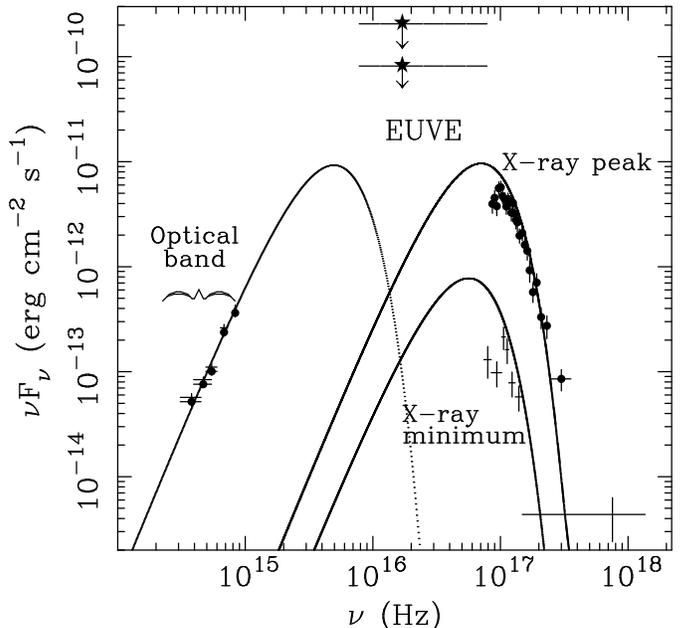

FIG. 6.— Broad-band energy spectrum of RX J0806.3+1527 as inferred from the *Chandra*, VLT and TNG (I02) observations and *EUVE* upper limits (inferred assuming $kT = 65$ eV and $N_H = 3$ and $5 \times 10^{20}$ cm$^{-2}$). Solid lines represent the unabsorbed blackbody models corresponding to the high and low states of the 321 s modulation (without the pile-up correction). The dotted line represents one of the possible fitting blackbody models for the optical band. Note that the optical measurements are a factor of about 100 higher than predicted by the X-ray blackbody.

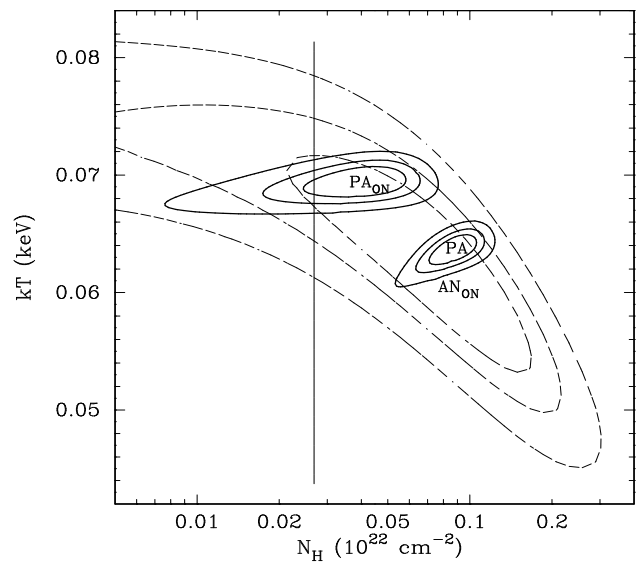

FIG. 7.— Confidence regions of the absorption $N_H$ as a function of the blackbody temperature $kT$ for the three different models described in the text. The vertical solid line marks the Galactic $N_H$ in the direction of the source ($2.7 \times 10^{20}$ cm$^{-2}$).

---

[1] Data were obtained by means of the EUVE Skymap count rate tool at the Center for EUV Astrophysics which is not available anymore. EUVE skymap data can be still obtained through the SkyView facility at http://skyview.gsfc.nasa.gov/cgi-bin/skvbasic.pl; see also Bowyer et al. 1991

flux in the EUV band. Note that such a value is a factor of about 20-200 smaller than that extrapolated for RX J1914.4+2456 by means of ROSAT data (Cropper et al. 1998). Future observations in the EUV band would be extremely important to better constrain the correct value for the total flux and the frequency of the energy peak(s) for both sources.

Under the hypothesis that the mass-donor star fills its Roche-lobe, its mass is $M_2 = 0.13 M_\odot$ (see I02). If the mass transfer rate $\dot{M}$ is driven by angular momentum losses through gravitational radiation, then the expected accretion luminosity is in the range $1–5 \times 10^{35}$ erg s$^{-1}$ (assuming efficient tidal coupling; see I02). This is inconsistent with the blackbody inferred form the X-ray data and marginally consistent with the possible bolometric luminosity. This, together with the detection of spin-up of the 321 s orbital period, argue against the Algol–like scenario.

Alternatively, the assumption that the secondary star fills its Roche-lobe might be inaccurate: in this case perhaps X-ray irradiation of the companion surface might trigger a wind causing mass transfer towards the primary white dwarf at a smaller rate (similar to the known case of LMXB with main sequence or white dwarf companion stars; Tavani & London 1993). Less likely, the source might lie outside the Galaxy (distance of the order of 2–3 kpc) and correspondingly more luminous. In the latter hypothesis the accreting source might be an eclipsing neutron star at a luminosity level of $\sim 10^{34}$ erg s$^{-1}$, even though the absence of conspicuous emission above 2 keV is difficult to account for in this scenario (not mentioning that the blackbody radius would increase to an uncomfortably high value of $\sim 100$ km). Note that the value of $N_H$ inferred from the X-ray spectrum is not useful, since local matter could be present around RX J0806.3+1527. Moreover, beyond about 500 pc there is no more neutral ISM to cause additional absorption. The latter considerations can be easily understood by looking at Figure 7, where we plot the confidence regions of the $N_H$ as a function of the blackbody temperature for three different cases: the phase-averaged (marked PA), the 0.5–0.9 phase interval (PA$_{ON}$), and the 0.5–0.9 phase interval of the annular region (AN$_{ON}$) spectra (note that the former two spectra were fitted taking into account the pile-up correction as described in the previous sections). The spectral parameters inferred from the AN$_{ON}$ spectrum are poorly constrained, and only an upper limit of $N_H < 3 \times 10^{21}$ cm$^{-2}$ ($3\sigma$) can be set. The other two regions have $N_H$ values that overlap, even though the temperatures are significantly different from each other (a higher temperature during the modulation-on phase interval). Note that the $N_H$ in the PA$_{ON}$ spectrum is consistent with the Galactic value in the direction of RX J0806.3+1527 ($\sim 2.7 \times 10^{20}$ cm$^{-2}$; Dickey & Lockman, 1990). On the other hand, we note that the $N_H$ value of the PA spectrum is significantly larger than the Galactic value, suggesting local absorption during the eclipse.

(iv) The over-abundance of He, C, O, and N seen in the VLT optical spectra of RX J0806.3+1527 are thought to reflect the chemical abundances of the secondary star. Therefore, the same elements are expected to be present also in the X-ray spectrum if matter is transferred bwteen the two stars. In this respect, the presence of complex edge features (even though their inclusion is not statistically significant) in the phase-averaged spectrum of RX J0806.3+1527 might be a potentially important finding for the study of the chemical abundances. The presence of such lines must be confirmed through the analysis of higher resolution X-ray spectra than those from ACIS–S, such as the *XMM-Newton* RGS and *Chandra* LETG.

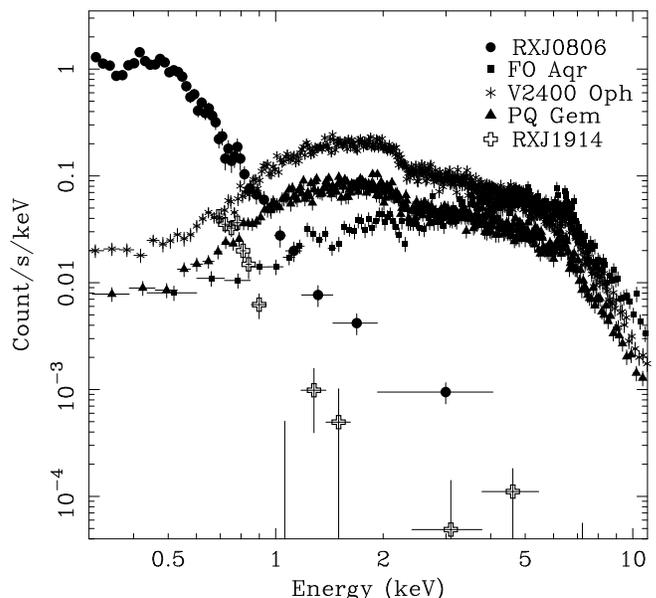

FIG. 8.— *ASCA* and *Chandra* spectral comparison of RX J0806.3+1527, RX J1914.4+2456, and a number of IPs. See the text for details.

Specifically, assuming that the absorption model is correct, we identified the relatively broad feature at 0.64 keV as the H-like N VII edge at 0.67 keV, while the 0.53 keV feature might be due to the H-like C VI edge and/or the H-like N VI edge at 0.49 keV and 0.55 keV, respectively. Finally, the 1.26 keV feature might be due to the H-like Ne IX edge at 1.20 keV. Due to their marginal statistical significance we prefer not to comment further on them, and to wait for higher statistics future observations.

(v) In order to check the IP scenario we compared the spectral X-ray emissions of RX J0806.3+1527 and RX J1914.4+2456 with a number of IPs that are thought to be stream-fed sources (FO Aqr, V2400 Oph and PQ Gem; Norton et al. 2002). Figure 8 shows the result of this comparison: all the data are taken from the *ASCA* database except RX J0806.3+1527 (for which we used *Chandra* data). Due to the different energy responses of *ASCA* and *Chandra*, only the 0.7–5 keV energy interval can be compared. It is evident how the RX J0806.3+1527 and RX J1914.4+2456 spectra stand out for being softer than the IP spectra. It is difficult to understand how such different spectra could be produced with different viewing angles. A similar result was also obtained using archival *ASCA* data of RX J0558.0+5353 and RE 0751+14, which are two of the softest IPs discovered by *ROSAT* (Haberl & Motch 1995).

(vi) Recently, a new double degenerate binary system with an orbital period of 620 s has been identified (KUV 01584−0939, alias Cet3; Warner & Woudt, 2002). Optical spectra show only strong He II lines from the Pickering series (Woudt & Warner 2003), classifying the source as a member of the AM CVn class. The source it is not included in the RASS catalogues, implying faint X-ray emission (if any). The orbital period of KUV 01584−0939 is very close to that of RX J1914.4+2456 (569 s). Correspondingly, if RX J0806.3+1527 and RX J1914.4+2456 belong to the AM CVn class, we must conclude that something happens in the accretion process (e.g., from disk to stream accretion) as the orbital period approaches 10 minutes (based on the inferred mass of the secondaries, RX J0806.3+1527 and RX J1914.4+2456 eventually evolve into AM CVn systems).

Alternatively, AM CVns and X-ray double degenerate systems (XDDBs) might come from two different evolutionary tracks.

In conclusion, the detected phase shift between the optical and X-ray light curves confirms the presence of the X-ray reprocessing mechanism in RX J0806.3+1527, likely a tight binary system hosting two degenerate stars. The X-ray underluminosity together with the recently detected $\dot{P}$ argue in favor of the unipolar inductor model or of a companion star not filling in its Roche lobe. Moreover, based on the *Chandra* spectrum and its comparison with those of other cataclysmic variables, we can resonably exclude that RX J0806.3+1527 is an intermediate polar.

Future observational studies (optical, IR and X-rays) are expected to distinguish among different scenarios. Among others the monitoring of the period derivative $\dot{P}$, optical phase resolved spectroscopy, and polarimetric studies will be especially important in this respect. Finally, the source represents one of the most promising targets for gravitational wave detection from binary motion, easily detectable by the LISA mission (Nelemans et al. 2001).

G.L.I. is deeply grateful to Paul Plucinsky of the *Chandra* Team and to Guenther Hasinger, Piero Rosati, Roberto Gilmozzi and Martino Romaniello for their help in the planning end execution of the *Chandra* and VLT observations. Support for this work was provided by the National Aeronautics and Space Administration through *Chandra* Award Number GO2-3084X issued by the *Chandra* X-ray Observatory Center, which is operated by the Smithsonian Astrophysical Observatory for and on behalf of the National Aeronautics Space Administration under contract NAS8-39073. C.W.M.'s contribution to this work was performed under the auspices of the U.S. Department of Energy by University of California Lawrence Livermore National Laboratory under contract No. W-7405-Eng-48. This work is supported through ASI, CNR and Ministero dell'Università e Ricerca Scientifica e Tecnologica (MURST-COFIN) grants.